\documentstyle[12pt]{article}

\begin{document}

\hoffset = -1truecm
\voffset = -2truecm

\title{\bf On a subtle point of sum rules calculations: toy model.
}

\author{{\bf A.A.Penin}\\
{\rm and}\\
{\bf A.A.Pivovarov}\\
{\it Institute for Nuclear  Research of the Russian Academy of Sciences},\\
{\it Moscow 117312, Russia}}
\date{}
\maketitle

\begin{abstract}
We consider a two-point correlator in massless $\phi^3$
model within the ladder approximation . The spectral
density of the correlator is known explicitly and does
not contain any resonances.  Meanwhile  making use of the
standard sum rules technique with a simple "resonance +
continuum" model of the spectrum allows to predict parameters
of the "resonance" very accurately in the sense that all
necessary criteria of stability are perfectly satisfied.
\end{abstract}

\vspace{0.5in}
PACS number(s): 11.50.Li

\thispagestyle{empty}

\newpage

$QCD$ sum rules have revealed themselves as a
powerful method to study hadron properties [1-4].
However certain effects were discovered that disprove
efficiency  of the method. Thus in the very beginning of
$QCD$ sum rules it has been found that instantons lead to
the appearance of so-called exponential terms which are
omitted within standard operator product expansion $(OPE)$
and induce the strong dependence of sum rules results on
detailed structure of $QCD$ vacuum \cite{3}.  Recently it
has been  pointed out that sum rules can also suffer from
infrared renormalons  that cannot be associated with
the expectation value of local operators \cite{4}.
Fortunately, in most of the practically interesting cases
these undesirable effects can be separated.  Though
some important questions still
remain unclear. The use of sum rules implies hadron
properties to be mainly determined by several leading terms
of asymptotic expansion of the correlator of relevant
interpolating currents in deep Euclidean domain that cannot
be guaranteed by itself.  This problem was studied  within
two-dimensional electrodynamic
(the Schwinger model) \cite{sm} where the exact solution is known
and sum rules calculations can be directly checked.  In
these papers the stability of the result with respect to
inclusion of higher order corrections was proposed as an
intrinsic criterion of sum rules applicability.  In the
present paper we demonstrate  within another exactly soluble
model an opposite example  when truncation of the asymptotic
expansion of a correlator leads to missing  the main
properties of its spectral density while the formal
stability requirement for corresponding sum rules is
completely satisfied .  This situation reflects some general
property of sum rules and can be realized in $QCD$ as well.

Let us describe the model.  We consider a ladder
approximation of massless $\phi^3$ model.
In the $\phi^3$ model the expansion
parameter is dimensionless ratio of the dimensionful
coupling constant and the energy of the process in question,
so the interaction dies out in the limit of
infinitely large energies. On the other hand at low energies
the parameter of the expansion becomes large and the model
requires a non-perturbative analysis.
Thus it would be instructive to
investigate the model along the line of ordinary $QCD$
methods. Not to be taken quite seriously it gives
nevertheless a way to go beyond the perturbation theory
because  here  the explicit expressions for diagrams in any
order of loop expansions are known [8-10]. The model
is attractive also because  other attempts to break the
bound of perturbation theory tend to be in dimensions
different from 4.

We modify
the usual $\phi^3$ model slightly  to make it more
convenient for our purpose.  The Lagrangian of our model
reads
$$
L=L_0+e_1\varphi^2 A +e_2 \phi^2 A \eqno (1)
$$
where $L_0$ is a free knetic term for all fields
and we  choose $e_2=-e_1=e$.
All questions of stability
of the model (the existence of a stable ground state,
for example) remain beyond the scope of
our toy consideration.

We study a correlator of two composite
operators $j=\varphi\phi$ in the ladder approximation.
The correlator has the form \cite{6}
$$
\Pi(q)=i\int\langle 0|{\rm T}j(x)j(0)|0\rangle e^{iqx}{\rm d}x,
{}~~\Pi(Q^2)={1\over 16\pi^2}\ln\left({\mu^2\over Q^2}
\right)+\Delta \Pi(Q^2)
\eqno (2)
$$
where
$$
\Delta \Pi(Q^2) ={1\over 16\pi^2}
\sum_{L=2}^{\infty}\left(-{e^2\over 16\pi^2Q^2}\right)^{L-1}
\left(\begin{array}{c} 2L\\ L
\end{array} \right)\zeta(2L-1),
\eqno (3)
$$
$Q^2=-q^2$ and $\zeta(z)$ is the Riemann's $\zeta$-function.

Since the coupling constant
$e$ is dimensionful the expansion~(3)
simulates power corrections or  $OPE$
of ordinary $QCD$. Setting $e^2/16\pi^2=1$ we get
$$
p(Q^2)\equiv 16\pi^2\Pi(Q^2)=\ln\left({\mu^2\over Q^2}\right)+\Delta p(Q^2)
$$
$$
\Delta p(Q^2) =\sum_{L=2}^{\infty}
\left(-{1\over Q^2}\right)^{L-1} \left(\begin{array}{c}
2L\\ L \end{array}
\right)\zeta(2L-1)
\eqno (4)
$$
in analogy with $QCD$ where the scale is given and all
condensates are expressed through $\Lambda_{QCD}$.

What we see first is the alternating character of the
series~(4) in Euclidean direction ($Q^2>0$) due
to the special choice of interaction~(1). This was
actually the reason to
choose all decorations for the simple $\phi^3$ model, in
which the series has the same sign
for each term for Euclidean $q$.

The sum~(4) has the following closed form \cite{7}
$$
-\sum_{n=1}^{\infty}nQ^2\left[\left(1+{4\over n^2Q^2}\right)^{-{1\over
2}}-1+{2\over n^2Q^2}\right]
$$
and does not contain any resonances.

The spectral density for the function
$\Delta p(Q^2)$
reads
$$
\Delta \rho(s)=-\sum_{n=1}^{\infty}
{ns\sqrt{s}\over\sqrt{{4\over n^2}-s}}\ \theta
\left({4\over n^2}-s\right)\theta (s).
\eqno (5)
$$
Unfortunately the whole spectral density of
the correlator, $\rho(s)=1+\Delta \rho(s)$, has negative sign
in some domains.  In case $e_1e_2>0$ it would have a
correct positive sign but the cut would be situated in the
wrong place of the complex plane (negative semiaxis).  So,
any choice of the interaction sign leads to unphysical
spectral density and the set of ladder diagrams is hardly
representative for the exact correlator $\langle 0|{\rm
T}j(x)j(0)|0\rangle$.
The same situation is realized in $QCD$.
If one takes seriously the leading order correlator for
vector currents, for example, one finds that the
spectral density contains an uphysical pole at
$Q^2=\Lambda_{QCD}^2$ and does not satisfy the
spectrality condition being negative at
$s<\Lambda_{QCD}^2$.

We omit these delicate points and proceed as in $QCD$.
We pose a question: does the asymptotic behavior of the
correlator~(2) contradict to the presence of a resonance
in the considered channel?
Namely, whether the expansion~(4) can be described
successfully with the simple formula
$$
\rho^{test}(s)=F\delta(s-m^2)+\theta(s-s_0)
\eqno (6)
$$
for the spectral density $\rho(s)$.

An explicit expansion for the correlator reads
$$
\Delta p(Q^2)=-{6 \zeta(3)\over Q^2}
+{20 \zeta(5)\over Q^4}-{70 \zeta(7)\over Q^6}
+{252 \zeta(9)\over Q^8}
+\ldots
\eqno (7)
$$
while the "test" form of the correlator is
$$
p^{test}(Q^2)=\ln\left({\mu^2\over Q^2+s_0}\right)+{F\over
Q^2+m^2}.
\eqno (8)
$$

We intend to connect the expressions~(7) and~(8) by means of
sum rules.
First we use finite energy sum rules $(FESR)$ \cite{8}
$$
\int_{0}^{s_0}s^k\rho(s){\rm d}s=
\int_{0}^{s_0}s^k\rho^{test}(s){\rm d}s
\eqno (9)
$$
where $k=0,~1,~2$ because the test spectral density has
three parameters to be determined. Eq.~(9) has an
explicit form
$$
F=s_0-6 \zeta(3), ~~Fm^2={1\over 2}s_0^2-20
\zeta(5), ~~Fm^4={1\over 3}s_0^3-70 \zeta(7)
\eqno (10)
$$
that leads to the equation for determination of
the duality threshold  $s_0$
$$
{1\over 12}s_0^4-2 \zeta(3) s_0^3+20 \zeta(5)
s_0^2 -70 \zeta(7) s_0 +420 \zeta(7) \zeta(3)-400
\zeta(5)^2=0.
\eqno (11)
$$
It has solutions $s_0=16.9$ and ${s'_0}=2.24$,
and for corresponding parameters $F=9.67$ and $m^2=12.6$
wlilst $F'=-4.97$ and $m'^2=3.67$.
We will study the first solution because the second one
gives an unnatural relation between "resonance mass" and
"duality interval" ${s'_0} < m'^2$.

Now we check corresponding expressions in the Borel
sum  rules approach  \cite{1}. The Borel
transformation $p(M^2)$ of the function $p(Q^2)$ is

$$
p(M^2)=1-{6 \zeta(3)\over M^2}
+{20 \zeta(5)\over M^4}-{35 \zeta(7)\over M^6}+{42 \zeta(9)\over M^8}
+\ldots,
\eqno (12)
$$
continuum gives
$$
c(M^2)={\rm exp}\left(-{s_0\over M^2}\right)
\eqno (13)
$$
and the resonance contribution reads
$$
r(M^2)={F\over M^2}{\rm exp}\left(-{m^2\over M^2}\right).
\eqno (14)
$$
We have plotted these functions in Fig.~1. As we see our
"test" representation  accurately simulates the asymptotic
form of the correlator at $M^2>8$.  Equating the functions
$p(M^2)$ and $r(M^2)+c(M^2)$ we obtain the Borel sum rules
to determine the parameters of the resonance.  The sum
rules look like ordinary $QCD$ sum rules.  Fig.~2 shows the
dependence of mass $m^2$ on the Borel variable $M^2$ with
the parameters $s_o$ and $F$ given by $FESR$ for the
different numbers of power corrections included.  For other
close values of the parameters $s_o$ and $F$ the results
are not very different from fig.~2.  A stable region is
reached for $8<M^2<18$ where, on the one hand, higher order
power corrections are convergent and, on the other hand,
the continuum contribution~(13) does not prevail over the
resonance one~(14).  The best stability is obtained at the
optimal values $s_o=16.6$, $F=9.39$ that shows a
consistency of the Borel sum rules and $FESR$. Furthermore,
from the curves in fig.~2 we see that the inclusion of the
high order power corrections does not destroy the Borel sum
rules and even extends the region of stability.

Let us emphasize that the sum rules are perfectly saturated
by the artificially introduced resonance and show very good
stability though the exact spectral density does not
contained any  resonance singularities.  The reason of this
phenomenon is quite transparent:  using sum rules we neglect
the high order terms of large momentum expansion which do
not affect the rough integral characteristics of the
spectral density but are essentially responsible for its
local behavior. The extreme sensitivity of the local form of
the spectral density  to the high orders of perturbative
expansion  can be easily demonstrated within considered
model. For example, substituting $\zeta$-functions for $L>2$
by units in the series~(4) one neglects the terms with $n>2$
in eq.~(5) and gets the modified spectral density

$$
\Delta \tilde\rho(s)=-s\sqrt{s}\ \theta (s)
\left({\theta(4-s)\over\sqrt{4-s}}
+{\theta(1-s)\over\sqrt{1-s}}\right).
\eqno (15)
$$
As we see the tiny variation of the coefficients of the
asymptotic expansion leads to a drastic change of the
spectral density at low scale $s<4/9$.  Indeed, the
modified spectral density has two  singular points at
$s=4~{\rm and}~1$ while the original expression has an
infinite number of singularities at  $s_n=4/n^2,~
n=1,2,\ldots$ At the same time  this variation does not
really affect the sum rules result that changes slightly
(less then $10\%$):  $ {\tilde s_0=17.9,}~~{\tilde
F=10.7,}~~{\tilde m^2=13.0}$.

The lesson we drew from the above consideration is that
applying sum rules technique one can find sum rules
to be stable while the leading terms of $OPE$ are hardly
related to the hadron parameters, {\it i.e.} the
non-perturbative information contained in the basic $QCD$
condensates is essentially insufficient to determine the
hadron properties or the physical spectral density has
unexpectedly complicated form.  At the same time a direct
method to distinguish and separate this effect is absent
now. So sum rules approach faces a subtle problem and
another criterion of reliability of sum rules predictions
is necessary instead of naive stability requirement.

\vspace{0.5cm}
\noindent
{\large \bf Acknowledgments}

\noindent
This work is supported in part
by Russian Fund for Fundamental Research under
Contract No.  93-02-14428, 94-02-06427 and by Soros
Foundation.

\vspace{0.5cm}
\noindent
{\large \bf Figure Captions}

\noindent
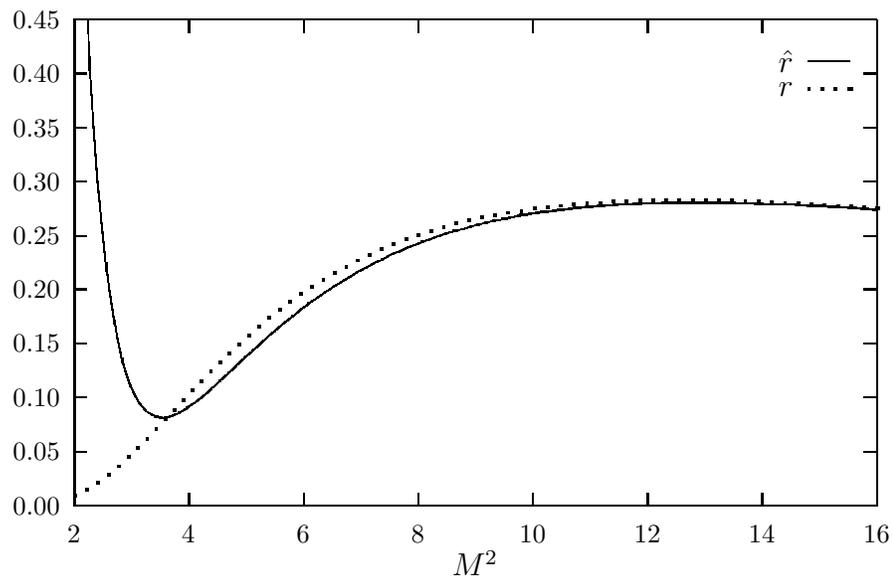
Fig. 1.  The resonance contribution $r(M^2)$ to the Borel
sum  rules (eq.(14)) and the functioin
$\hat r(M^2)=p(M^2)-c(M^2)$ (eq.(12,13)).

\noindent
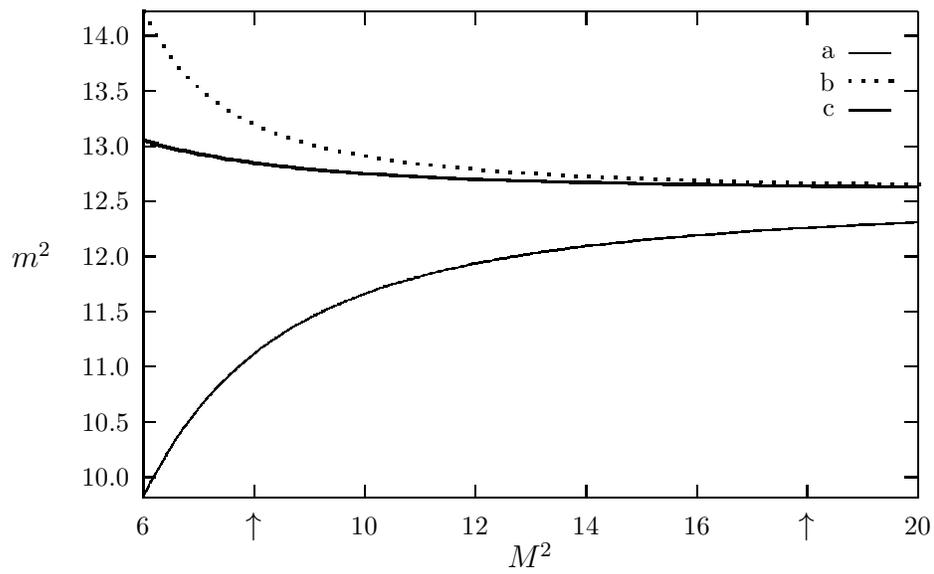
Fig. 2.  The mass $m^2$  plotted as a function of
Borel variable $M^2$ with two (a), three (b) and four (c)
orders of power corrections included. The arrows mark
stability interval of the Borel sum  rules.

\newpage
\vspace*{5cm}

\setlength{\unitlength}{0.240900pt}
\ifx\plotpoint\undefined\newsavebox{\plotpoint}\fi
\sbox{\plotpoint}{\rule[-0.200pt]{0.400pt}{0.400pt}}%
\begin{picture}(1500,900)(0,0)
\font\gnuplot=cmr10 at 10pt
\gnuplot
\sbox{\plotpoint}{\rule[-0.200pt]{0.400pt}{0.400pt}}%
\put(176.0,113.0){\rule[-0.200pt]{303.534pt}{0.400pt}}
\put(176.0,113.0){\rule[-0.200pt]{4.818pt}{0.400pt}}
\put(154,113){\makebox(0,0)[r]{0.00}}
\put(1416.0,113.0){\rule[-0.200pt]{4.818pt}{0.400pt}}
\put(176.0,198.0){\rule[-0.200pt]{4.818pt}{0.400pt}}
\put(154,198){\makebox(0,0)[r]{0.05}}
\put(1416.0,198.0){\rule[-0.200pt]{4.818pt}{0.400pt}}
\put(176.0,283.0){\rule[-0.200pt]{4.818pt}{0.400pt}}
\put(154,283){\makebox(0,0)[r]{0.10}}
\put(1416.0,283.0){\rule[-0.200pt]{4.818pt}{0.400pt}}
\put(176.0,368.0){\rule[-0.200pt]{4.818pt}{0.400pt}}
\put(154,368){\makebox(0,0)[r]{0.15}}
\put(1416.0,368.0){\rule[-0.200pt]{4.818pt}{0.400pt}}
\put(176.0,453.0){\rule[-0.200pt]{4.818pt}{0.400pt}}
\put(154,453){\makebox(0,0)[r]{0.20}}
\put(1416.0,453.0){\rule[-0.200pt]{4.818pt}{0.400pt}}
\put(176.0,537.0){\rule[-0.200pt]{4.818pt}{0.400pt}}
\put(154,537){\makebox(0,0)[r]{0.25}}
\put(1416.0,537.0){\rule[-0.200pt]{4.818pt}{0.400pt}}
\put(176.0,622.0){\rule[-0.200pt]{4.818pt}{0.400pt}}
\put(154,622){\makebox(0,0)[r]{0.30}}
\put(1416.0,622.0){\rule[-0.200pt]{4.818pt}{0.400pt}}
\put(176.0,707.0){\rule[-0.200pt]{4.818pt}{0.400pt}}
\put(154,707){\makebox(0,0)[r]{0.35}}
\put(1416.0,707.0){\rule[-0.200pt]{4.818pt}{0.400pt}}
\put(176.0,792.0){\rule[-0.200pt]{4.818pt}{0.400pt}}
\put(154,792){\makebox(0,0)[r]{0.40}}
\put(1416.0,792.0){\rule[-0.200pt]{4.818pt}{0.400pt}}
\put(176.0,877.0){\rule[-0.200pt]{4.818pt}{0.400pt}}
\put(154,877){\makebox(0,0)[r]{0.45}}
\put(1416.0,877.0){\rule[-0.200pt]{4.818pt}{0.400pt}}
\put(176.0,113.0){\rule[-0.200pt]{0.400pt}{4.818pt}}
\put(176,68){\makebox(0,0){2}}
\put(176.0,857.0){\rule[-0.200pt]{0.400pt}{4.818pt}}
\put(356.0,113.0){\rule[-0.200pt]{0.400pt}{4.818pt}}
\put(356,68){\makebox(0,0){4}}
\put(356.0,857.0){\rule[-0.200pt]{0.400pt}{4.818pt}}
\put(536.0,113.0){\rule[-0.200pt]{0.400pt}{4.818pt}}
\put(536,68){\makebox(0,0){6}}
\put(536.0,857.0){\rule[-0.200pt]{0.400pt}{4.818pt}}
\put(716.0,113.0){\rule[-0.200pt]{0.400pt}{4.818pt}}
\put(716,68){\makebox(0,0){8}}
\put(716.0,857.0){\rule[-0.200pt]{0.400pt}{4.818pt}}
\put(896.0,113.0){\rule[-0.200pt]{0.400pt}{4.818pt}}
\put(896,68){\makebox(0,0){10}}
\put(896.0,857.0){\rule[-0.200pt]{0.400pt}{4.818pt}}
\put(1076.0,113.0){\rule[-0.200pt]{0.400pt}{4.818pt}}
\put(1076,68){\makebox(0,0){12}}
\put(1076.0,857.0){\rule[-0.200pt]{0.400pt}{4.818pt}}
\put(1256.0,113.0){\rule[-0.200pt]{0.400pt}{4.818pt}}
\put(1256,68){\makebox(0,0){14}}
\put(1256.0,857.0){\rule[-0.200pt]{0.400pt}{4.818pt}}
\put(1436.0,113.0){\rule[-0.200pt]{0.400pt}{4.818pt}}
\put(1436,68){\makebox(0,0){16}}
\put(1436.0,857.0){\rule[-0.200pt]{0.400pt}{4.818pt}}
\put(176.0,113.0){\rule[-0.200pt]{303.534pt}{0.400pt}}
\put(1436.0,113.0){\rule[-0.200pt]{0.400pt}{184.048pt}}
\put(176.0,877.0){\rule[-0.200pt]{303.534pt}{0.400pt}}
\put(806,23){\makebox(0,0){$M^2$}}
\put(176.0,113.0){\rule[-0.200pt]{0.400pt}{184.048pt}}
\put(1306,812){\makebox(0,0)[r]{$\hat r$}}
\multiput(198.61,831.75)(0.447,-17.877){3}{\rule{0.108pt}{10.900pt}}
\multiput(197.17,854.38)(3.000,-58.377){2}{\rule{0.400pt}{5.450pt}}
\multiput(201.58,772.08)(0.493,-7.277){23}{\rule{0.119pt}{5.762pt}}
\multiput(200.17,784.04)(13.000,-172.042){2}{\rule{0.400pt}{2.881pt}}
\multiput(214.58,595.49)(0.493,-4.977){23}{\rule{0.119pt}{3.977pt}}
\multiput(213.17,603.75)(13.000,-117.746){2}{\rule{0.400pt}{1.988pt}}
\multiput(227.58,474.47)(0.493,-3.431){23}{\rule{0.119pt}{2.777pt}}
\multiput(226.17,480.24)(13.000,-81.236){2}{\rule{0.400pt}{1.388pt}}
\multiput(240.58,390.42)(0.492,-2.521){21}{\rule{0.119pt}{2.067pt}}
\multiput(239.17,394.71)(12.000,-54.711){2}{\rule{0.400pt}{1.033pt}}
\multiput(252.58,334.48)(0.493,-1.567){23}{\rule{0.119pt}{1.331pt}}
\multiput(251.17,337.24)(13.000,-37.238){2}{\rule{0.400pt}{0.665pt}}
\multiput(265.58,296.39)(0.493,-0.972){23}{\rule{0.119pt}{0.869pt}}
\multiput(264.17,298.20)(13.000,-23.196){2}{\rule{0.400pt}{0.435pt}}
\multiput(278.58,272.67)(0.493,-0.576){23}{\rule{0.119pt}{0.562pt}}
\multiput(277.17,273.83)(13.000,-13.834){2}{\rule{0.400pt}{0.281pt}}
\multiput(291.00,258.93)(0.874,-0.485){11}{\rule{0.786pt}{0.117pt}}
\multiput(291.00,259.17)(10.369,-7.000){2}{\rule{0.393pt}{0.400pt}}
\put(303,251.17){\rule{2.700pt}{0.400pt}}
\multiput(303.00,252.17)(7.396,-2.000){2}{\rule{1.350pt}{0.400pt}}
\multiput(316.00,251.61)(2.695,0.447){3}{\rule{1.833pt}{0.108pt}}
\multiput(316.00,250.17)(9.195,3.000){2}{\rule{0.917pt}{0.400pt}}
\multiput(329.00,254.59)(1.267,0.477){7}{\rule{1.060pt}{0.115pt}}
\multiput(329.00,253.17)(9.800,5.000){2}{\rule{0.530pt}{0.400pt}}
\multiput(341.00,259.59)(0.824,0.488){13}{\rule{0.750pt}{0.117pt}}
\multiput(341.00,258.17)(11.443,8.000){2}{\rule{0.375pt}{0.400pt}}
\multiput(354.00,267.59)(0.728,0.489){15}{\rule{0.678pt}{0.118pt}}
\multiput(354.00,266.17)(11.593,9.000){2}{\rule{0.339pt}{0.400pt}}
\multiput(367.00,276.58)(0.652,0.491){17}{\rule{0.620pt}{0.118pt}}
\multiput(367.00,275.17)(11.713,10.000){2}{\rule{0.310pt}{0.400pt}}
\multiput(380.00,286.58)(0.543,0.492){19}{\rule{0.536pt}{0.118pt}}
\multiput(380.00,285.17)(10.887,11.000){2}{\rule{0.268pt}{0.400pt}}
\multiput(392.00,297.58)(0.539,0.492){21}{\rule{0.533pt}{0.119pt}}
\multiput(392.00,296.17)(11.893,12.000){2}{\rule{0.267pt}{0.400pt}}
\multiput(405.00,309.58)(0.539,0.492){21}{\rule{0.533pt}{0.119pt}}
\multiput(405.00,308.17)(11.893,12.000){2}{\rule{0.267pt}{0.400pt}}
\multiput(418.00,321.58)(0.539,0.492){21}{\rule{0.533pt}{0.119pt}}
\multiput(418.00,320.17)(11.893,12.000){2}{\rule{0.267pt}{0.400pt}}
\multiput(431.00,333.58)(0.496,0.492){21}{\rule{0.500pt}{0.119pt}}
\multiput(431.00,332.17)(10.962,12.000){2}{\rule{0.250pt}{0.400pt}}
\multiput(443.00,345.58)(0.590,0.492){19}{\rule{0.573pt}{0.118pt}}
\multiput(443.00,344.17)(11.811,11.000){2}{\rule{0.286pt}{0.400pt}}
\multiput(456.00,356.58)(0.539,0.492){21}{\rule{0.533pt}{0.119pt}}
\multiput(456.00,355.17)(11.893,12.000){2}{\rule{0.267pt}{0.400pt}}
\multiput(469.00,368.58)(0.543,0.492){19}{\rule{0.536pt}{0.118pt}}
\multiput(469.00,367.17)(10.887,11.000){2}{\rule{0.268pt}{0.400pt}}
\multiput(481.00,379.58)(0.590,0.492){19}{\rule{0.573pt}{0.118pt}}
\multiput(481.00,378.17)(11.811,11.000){2}{\rule{0.286pt}{0.400pt}}
\multiput(494.00,390.58)(0.590,0.492){19}{\rule{0.573pt}{0.118pt}}
\multiput(494.00,389.17)(11.811,11.000){2}{\rule{0.286pt}{0.400pt}}
\multiput(507.00,401.58)(0.652,0.491){17}{\rule{0.620pt}{0.118pt}}
\multiput(507.00,400.17)(11.713,10.000){2}{\rule{0.310pt}{0.400pt}}
\multiput(520.00,411.58)(0.600,0.491){17}{\rule{0.580pt}{0.118pt}}
\multiput(520.00,410.17)(10.796,10.000){2}{\rule{0.290pt}{0.400pt}}
\multiput(532.00,421.58)(0.652,0.491){17}{\rule{0.620pt}{0.118pt}}
\multiput(532.00,420.17)(11.713,10.000){2}{\rule{0.310pt}{0.400pt}}
\multiput(545.00,431.59)(0.728,0.489){15}{\rule{0.678pt}{0.118pt}}
\multiput(545.00,430.17)(11.593,9.000){2}{\rule{0.339pt}{0.400pt}}
\multiput(558.00,440.59)(0.728,0.489){15}{\rule{0.678pt}{0.118pt}}
\multiput(558.00,439.17)(11.593,9.000){2}{\rule{0.339pt}{0.400pt}}
\multiput(571.00,449.59)(0.758,0.488){13}{\rule{0.700pt}{0.117pt}}
\multiput(571.00,448.17)(10.547,8.000){2}{\rule{0.350pt}{0.400pt}}
\multiput(583.00,457.59)(0.824,0.488){13}{\rule{0.750pt}{0.117pt}}
\multiput(583.00,456.17)(11.443,8.000){2}{\rule{0.375pt}{0.400pt}}
\multiput(596.00,465.59)(0.824,0.488){13}{\rule{0.750pt}{0.117pt}}
\multiput(596.00,464.17)(11.443,8.000){2}{\rule{0.375pt}{0.400pt}}
\multiput(609.00,473.59)(0.874,0.485){11}{\rule{0.786pt}{0.117pt}}
\multiput(609.00,472.17)(10.369,7.000){2}{\rule{0.393pt}{0.400pt}}
\multiput(621.00,480.59)(0.950,0.485){11}{\rule{0.843pt}{0.117pt}}
\multiput(621.00,479.17)(11.251,7.000){2}{\rule{0.421pt}{0.400pt}}
\multiput(634.00,487.59)(0.950,0.485){11}{\rule{0.843pt}{0.117pt}}
\multiput(634.00,486.17)(11.251,7.000){2}{\rule{0.421pt}{0.400pt}}
\multiput(647.00,494.59)(1.123,0.482){9}{\rule{0.967pt}{0.116pt}}
\multiput(647.00,493.17)(10.994,6.000){2}{\rule{0.483pt}{0.400pt}}
\multiput(660.00,500.59)(0.874,0.485){11}{\rule{0.786pt}{0.117pt}}
\multiput(660.00,499.17)(10.369,7.000){2}{\rule{0.393pt}{0.400pt}}
\multiput(672.00,507.59)(1.378,0.477){7}{\rule{1.140pt}{0.115pt}}
\multiput(672.00,506.17)(10.634,5.000){2}{\rule{0.570pt}{0.400pt}}
\multiput(685.00,512.59)(1.123,0.482){9}{\rule{0.967pt}{0.116pt}}
\multiput(685.00,511.17)(10.994,6.000){2}{\rule{0.483pt}{0.400pt}}
\multiput(698.00,518.59)(1.378,0.477){7}{\rule{1.140pt}{0.115pt}}
\multiput(698.00,517.17)(10.634,5.000){2}{\rule{0.570pt}{0.400pt}}
\multiput(711.00,523.59)(1.267,0.477){7}{\rule{1.060pt}{0.115pt}}
\multiput(711.00,522.17)(9.800,5.000){2}{\rule{0.530pt}{0.400pt}}
\multiput(723.00,528.60)(1.797,0.468){5}{\rule{1.400pt}{0.113pt}}
\multiput(723.00,527.17)(10.094,4.000){2}{\rule{0.700pt}{0.400pt}}
\multiput(736.00,532.59)(1.378,0.477){7}{\rule{1.140pt}{0.115pt}}
\multiput(736.00,531.17)(10.634,5.000){2}{\rule{0.570pt}{0.400pt}}
\multiput(749.00,537.60)(1.651,0.468){5}{\rule{1.300pt}{0.113pt}}
\multiput(749.00,536.17)(9.302,4.000){2}{\rule{0.650pt}{0.400pt}}
\multiput(761.00,541.60)(1.797,0.468){5}{\rule{1.400pt}{0.113pt}}
\multiput(761.00,540.17)(10.094,4.000){2}{\rule{0.700pt}{0.400pt}}
\multiput(774.00,545.61)(2.695,0.447){3}{\rule{1.833pt}{0.108pt}}
\multiput(774.00,544.17)(9.195,3.000){2}{\rule{0.917pt}{0.400pt}}
\multiput(787.00,548.60)(1.797,0.468){5}{\rule{1.400pt}{0.113pt}}
\multiput(787.00,547.17)(10.094,4.000){2}{\rule{0.700pt}{0.400pt}}
\multiput(800.00,552.61)(2.472,0.447){3}{\rule{1.700pt}{0.108pt}}
\multiput(800.00,551.17)(8.472,3.000){2}{\rule{0.850pt}{0.400pt}}
\multiput(812.00,555.61)(2.695,0.447){3}{\rule{1.833pt}{0.108pt}}
\multiput(812.00,554.17)(9.195,3.000){2}{\rule{0.917pt}{0.400pt}}
\multiput(825.00,558.61)(2.695,0.447){3}{\rule{1.833pt}{0.108pt}}
\multiput(825.00,557.17)(9.195,3.000){2}{\rule{0.917pt}{0.400pt}}
\multiput(838.00,561.61)(2.695,0.447){3}{\rule{1.833pt}{0.108pt}}
\multiput(838.00,560.17)(9.195,3.000){2}{\rule{0.917pt}{0.400pt}}
\put(851,564.17){\rule{2.500pt}{0.400pt}}
\multiput(851.00,563.17)(6.811,2.000){2}{\rule{1.250pt}{0.400pt}}
\multiput(863.00,566.61)(2.695,0.447){3}{\rule{1.833pt}{0.108pt}}
\multiput(863.00,565.17)(9.195,3.000){2}{\rule{0.917pt}{0.400pt}}
\put(876,569.17){\rule{2.700pt}{0.400pt}}
\multiput(876.00,568.17)(7.396,2.000){2}{\rule{1.350pt}{0.400pt}}
\put(889,571.17){\rule{2.500pt}{0.400pt}}
\multiput(889.00,570.17)(6.811,2.000){2}{\rule{1.250pt}{0.400pt}}
\put(901,573.17){\rule{2.700pt}{0.400pt}}
\multiput(901.00,572.17)(7.396,2.000){2}{\rule{1.350pt}{0.400pt}}
\put(914,574.67){\rule{3.132pt}{0.400pt}}
\multiput(914.00,574.17)(6.500,1.000){2}{\rule{1.566pt}{0.400pt}}
\put(927,576.17){\rule{2.700pt}{0.400pt}}
\multiput(927.00,575.17)(7.396,2.000){2}{\rule{1.350pt}{0.400pt}}
\put(940,577.67){\rule{2.891pt}{0.400pt}}
\multiput(940.00,577.17)(6.000,1.000){2}{\rule{1.445pt}{0.400pt}}
\put(952,579.17){\rule{2.700pt}{0.400pt}}
\multiput(952.00,578.17)(7.396,2.000){2}{\rule{1.350pt}{0.400pt}}
\put(965,580.67){\rule{3.132pt}{0.400pt}}
\multiput(965.00,580.17)(6.500,1.000){2}{\rule{1.566pt}{0.400pt}}
\put(978,581.67){\rule{3.132pt}{0.400pt}}
\multiput(978.00,581.17)(6.500,1.000){2}{\rule{1.566pt}{0.400pt}}
\put(991,582.67){\rule{2.891pt}{0.400pt}}
\multiput(991.00,582.17)(6.000,1.000){2}{\rule{1.445pt}{0.400pt}}
\put(1003,583.67){\rule{3.132pt}{0.400pt}}
\multiput(1003.00,583.17)(6.500,1.000){2}{\rule{1.566pt}{0.400pt}}
\put(1016,584.67){\rule{3.132pt}{0.400pt}}
\multiput(1016.00,584.17)(6.500,1.000){2}{\rule{1.566pt}{0.400pt}}
\put(1328.0,812.0){\rule[-0.200pt]{15.899pt}{0.400pt}}
\put(1041,585.67){\rule{3.132pt}{0.400pt}}
\multiput(1041.00,585.17)(6.500,1.000){2}{\rule{1.566pt}{0.400pt}}
\put(1054,586.67){\rule{3.132pt}{0.400pt}}
\multiput(1054.00,586.17)(6.500,1.000){2}{\rule{1.566pt}{0.400pt}}
\put(1029.0,586.0){\rule[-0.200pt]{2.891pt}{0.400pt}}
\put(1092,587.67){\rule{3.132pt}{0.400pt}}
\multiput(1092.00,587.17)(6.500,1.000){2}{\rule{1.566pt}{0.400pt}}
\put(1067.0,588.0){\rule[-0.200pt]{6.022pt}{0.400pt}}
\put(1207,587.67){\rule{3.132pt}{0.400pt}}
\multiput(1207.00,588.17)(6.500,-1.000){2}{\rule{1.566pt}{0.400pt}}
\put(1105.0,589.0){\rule[-0.200pt]{24.572pt}{0.400pt}}
\put(1245,586.67){\rule{3.132pt}{0.400pt}}
\multiput(1245.00,587.17)(6.500,-1.000){2}{\rule{1.566pt}{0.400pt}}
\put(1220.0,588.0){\rule[-0.200pt]{6.022pt}{0.400pt}}
\put(1283,585.67){\rule{3.132pt}{0.400pt}}
\multiput(1283.00,586.17)(6.500,-1.000){2}{\rule{1.566pt}{0.400pt}}
\put(1258.0,587.0){\rule[-0.200pt]{6.022pt}{0.400pt}}
\put(1309,584.67){\rule{2.891pt}{0.400pt}}
\multiput(1309.00,585.17)(6.000,-1.000){2}{\rule{1.445pt}{0.400pt}}
\put(1321,583.67){\rule{3.132pt}{0.400pt}}
\multiput(1321.00,584.17)(6.500,-1.000){2}{\rule{1.566pt}{0.400pt}}
\put(1296.0,586.0){\rule[-0.200pt]{3.132pt}{0.400pt}}
\put(1347,582.67){\rule{3.132pt}{0.400pt}}
\multiput(1347.00,583.17)(6.500,-1.000){2}{\rule{1.566pt}{0.400pt}}
\put(1360,581.67){\rule{2.891pt}{0.400pt}}
\multiput(1360.00,582.17)(6.000,-1.000){2}{\rule{1.445pt}{0.400pt}}
\put(1334.0,584.0){\rule[-0.200pt]{3.132pt}{0.400pt}}
\put(1385,580.67){\rule{3.132pt}{0.400pt}}
\multiput(1385.00,581.17)(6.500,-1.000){2}{\rule{1.566pt}{0.400pt}}
\put(1398,579.67){\rule{3.132pt}{0.400pt}}
\multiput(1398.00,580.17)(6.500,-1.000){2}{\rule{1.566pt}{0.400pt}}
\put(1411,578.67){\rule{2.891pt}{0.400pt}}
\multiput(1411.00,579.17)(6.000,-1.000){2}{\rule{1.445pt}{0.400pt}}
\put(1423,577.67){\rule{3.132pt}{0.400pt}}
\multiput(1423.00,578.17)(6.500,-1.000){2}{\rule{1.566pt}{0.400pt}}
\put(1372.0,582.0){\rule[-0.200pt]{3.132pt}{0.400pt}}
\sbox{\plotpoint}{\rule[-0.500pt]{1.000pt}{1.000pt}}%
\put(1306,767){\makebox(0,0)[r]{$r$}}
\multiput(1328,767)(20.756,0.000){4}{\usebox{\plotpoint}}
\put(1394,767){\usebox{\plotpoint}}
\put(176,128){\usebox{\plotpoint}}
\put(176.00,128.00){\usebox{\plotpoint}}
\put(194.36,137.57){\usebox{\plotpoint}}
\put(211.50,149.27){\usebox{\plotpoint}}
\multiput(214,151)(17.065,11.814){0}{\usebox{\plotpoint}}
\put(228.45,161.23){\usebox{\plotpoint}}
\put(244.15,174.80){\usebox{\plotpoint}}
\put(259.43,188.86){\usebox{\plotpoint}}
\put(274.31,203.31){\usebox{\plotpoint}}
\put(288.99,217.99){\usebox{\plotpoint}}
\multiput(291,220)(14.078,15.251){0}{\usebox{\plotpoint}}
\put(303.15,233.16){\usebox{\plotpoint}}
\put(317.32,248.32){\usebox{\plotpoint}}
\put(331.76,263.22){\usebox{\plotpoint}}
\put(345.64,278.64){\usebox{\plotpoint}}
\put(360.08,293.54){\usebox{\plotpoint}}
\put(374.48,308.48){\usebox{\plotpoint}}
\put(388.78,323.52){\usebox{\plotpoint}}
\put(403.77,337.86){\usebox{\plotpoint}}
\multiput(405,339)(14.676,14.676){0}{\usebox{\plotpoint}}
\put(418.51,352.47){\usebox{\plotpoint}}
\put(433.77,366.54){\usebox{\plotpoint}}
\put(449.28,380.32){\usebox{\plotpoint}}
\put(465.13,393.72){\usebox{\plotpoint}}
\put(480.56,407.60){\usebox{\plotpoint}}
\multiput(481,408)(16.451,12.655){0}{\usebox{\plotpoint}}
\put(496.98,420.29){\usebox{\plotpoint}}
\put(513.67,432.62){\usebox{\plotpoint}}
\put(530.45,444.84){\usebox{\plotpoint}}
\multiput(532,446)(17.065,11.814){0}{\usebox{\plotpoint}}
\put(547.56,456.57){\usebox{\plotpoint}}
\put(565.23,467.45){\usebox{\plotpoint}}
\multiput(571,471)(17.928,10.458){0}{\usebox{\plotpoint}}
\put(583.08,478.04){\usebox{\plotpoint}}
\put(601.36,487.88){\usebox{\plotpoint}}
\put(619.43,498.08){\usebox{\plotpoint}}
\multiput(621,499)(18.845,8.698){0}{\usebox{\plotpoint}}
\put(638.19,506.94){\usebox{\plotpoint}}
\put(657.04,515.63){\usebox{\plotpoint}}
\multiput(660,517)(19.159,7.983){0}{\usebox{\plotpoint}}
\put(676.19,523.61){\usebox{\plotpoint}}
\put(695.57,531.06){\usebox{\plotpoint}}
\multiput(698,532)(19.838,6.104){0}{\usebox{\plotpoint}}
\put(715.20,537.75){\usebox{\plotpoint}}
\put(734.76,544.62){\usebox{\plotpoint}}
\multiput(736,545)(19.838,6.104){0}{\usebox{\plotpoint}}
\put(754.68,550.42){\usebox{\plotpoint}}
\multiput(761,552)(19.838,6.104){0}{\usebox{\plotpoint}}
\put(774.62,556.14){\usebox{\plotpoint}}
\put(794.85,560.81){\usebox{\plotpoint}}
\multiput(800,562)(20.136,5.034){0}{\usebox{\plotpoint}}
\put(815.06,565.47){\usebox{\plotpoint}}
\put(835.43,569.41){\usebox{\plotpoint}}
\multiput(838,570)(20.514,3.156){0}{\usebox{\plotpoint}}
\put(855.89,572.82){\usebox{\plotpoint}}
\multiput(863,574)(20.514,3.156){0}{\usebox{\plotpoint}}
\put(876.39,576.06){\usebox{\plotpoint}}
\put(896.89,579.32){\usebox{\plotpoint}}
\multiput(901,580)(20.694,1.592){0}{\usebox{\plotpoint}}
\put(917.51,581.54){\usebox{\plotpoint}}
\put(938.12,583.86){\usebox{\plotpoint}}
\multiput(940,584)(20.684,1.724){0}{\usebox{\plotpoint}}
\put(958.75,586.04){\usebox{\plotpoint}}
\multiput(965,587)(20.694,1.592){0}{\usebox{\plotpoint}}
\put(979.40,588.00){\usebox{\plotpoint}}
\put(1000.12,588.76){\usebox{\plotpoint}}
\multiput(1003,589)(20.694,1.592){0}{\usebox{\plotpoint}}
\put(1020.81,590.37){\usebox{\plotpoint}}
\multiput(1029,591)(20.756,0.000){0}{\usebox{\plotpoint}}
\put(1041.54,591.04){\usebox{\plotpoint}}
\put(1062.26,592.00){\usebox{\plotpoint}}
\multiput(1067,592)(20.756,0.000){0}{\usebox{\plotpoint}}
\put(1083.02,592.00){\usebox{\plotpoint}}
\put(1103.74,592.90){\usebox{\plotpoint}}
\multiput(1105,593)(20.756,0.000){0}{\usebox{\plotpoint}}
\put(1124.49,593.00){\usebox{\plotpoint}}
\multiput(1131,593)(20.756,0.000){0}{\usebox{\plotpoint}}
\put(1145.24,593.00){\usebox{\plotpoint}}
\put(1165.97,592.23){\usebox{\plotpoint}}
\multiput(1169,592)(20.756,0.000){0}{\usebox{\plotpoint}}
\put(1186.72,592.00){\usebox{\plotpoint}}
\multiput(1194,592)(20.756,0.000){0}{\usebox{\plotpoint}}
\put(1207.47,591.96){\usebox{\plotpoint}}
\put(1228.19,591.00){\usebox{\plotpoint}}
\multiput(1232,591)(20.756,0.000){0}{\usebox{\plotpoint}}
\put(1248.93,590.70){\usebox{\plotpoint}}
\put(1269.66,590.00){\usebox{\plotpoint}}
\multiput(1271,590)(20.684,-1.724){0}{\usebox{\plotpoint}}
\put(1290.35,588.43){\usebox{\plotpoint}}
\multiput(1296,588)(20.756,0.000){0}{\usebox{\plotpoint}}
\put(1311.09,587.83){\usebox{\plotpoint}}
\put(1331.77,586.17){\usebox{\plotpoint}}
\multiput(1334,586)(20.756,0.000){0}{\usebox{\plotpoint}}
\put(1352.51,585.58){\usebox{\plotpoint}}
\multiput(1360,585)(20.684,-1.724){0}{\usebox{\plotpoint}}
\put(1373.20,583.91){\usebox{\plotpoint}}
\put(1393.92,583.00){\usebox{\plotpoint}}
\multiput(1398,583)(20.694,-1.592){0}{\usebox{\plotpoint}}
\put(1414.62,581.70){\usebox{\plotpoint}}
\put(1435.31,580.05){\usebox{\plotpoint}}
\put(1436,580){\usebox{\plotpoint}}
\end{picture}

\begin{center}
{\large \bf Fig. 1}
\end{center}

\newpage
\vspace*{5cm}

\setlength{\unitlength}{0.240900pt}
\ifx\plotpoint\undefined\newsavebox{\plotpoint}\fi
\sbox{\plotpoint}{\rule[-0.200pt]{0.400pt}{0.400pt}}%
\begin{picture}(1500,900)(0,0)
\font\gnuplot=cmr10 at 10pt
\gnuplot
\sbox{\plotpoint}{\rule[-0.200pt]{0.400pt}{0.400pt}}%
\put(220.0,145.0){\rule[-0.200pt]{4.818pt}{0.400pt}}
\put(198,145){\makebox(0,0)[r]{10.0}}
\put(1416.0,145.0){\rule[-0.200pt]{4.818pt}{0.400pt}}
\put(220.0,232.0){\rule[-0.200pt]{4.818pt}{0.400pt}}
\put(198,232){\makebox(0,0)[r]{10.5}}
\put(1416.0,232.0){\rule[-0.200pt]{4.818pt}{0.400pt}}
\put(220.0,319.0){\rule[-0.200pt]{4.818pt}{0.400pt}}
\put(198,319){\makebox(0,0)[r]{11.0}}
\put(1416.0,319.0){\rule[-0.200pt]{4.818pt}{0.400pt}}
\put(220.0,405.0){\rule[-0.200pt]{4.818pt}{0.400pt}}
\put(198,405){\makebox(0,0)[r]{11.5}}
\put(1416.0,405.0){\rule[-0.200pt]{4.818pt}{0.400pt}}
\put(220.0,492.0){\rule[-0.200pt]{4.818pt}{0.400pt}}
\put(198,492){\makebox(0,0)[r]{12.0}}
\put(1416.0,492.0){\rule[-0.200pt]{4.818pt}{0.400pt}}
\put(220.0,579.0){\rule[-0.200pt]{4.818pt}{0.400pt}}
\put(198,579){\makebox(0,0)[r]{12.5}}
\put(1416.0,579.0){\rule[-0.200pt]{4.818pt}{0.400pt}}
\put(220.0,665.0){\rule[-0.200pt]{4.818pt}{0.400pt}}
\put(198,665){\makebox(0,0)[r]{13.0}}
\put(1416.0,665.0){\rule[-0.200pt]{4.818pt}{0.400pt}}
\put(220.0,752.0){\rule[-0.200pt]{4.818pt}{0.400pt}}
\put(198,752){\makebox(0,0)[r]{13.5}}
\put(1416.0,752.0){\rule[-0.200pt]{4.818pt}{0.400pt}}
\put(220.0,839.0){\rule[-0.200pt]{4.818pt}{0.400pt}}
\put(198,839){\makebox(0,0)[r]{14.0}}
\put(1416.0,839.0){\rule[-0.200pt]{4.818pt}{0.400pt}}
\put(220.0,113.0){\rule[-0.200pt]{0.400pt}{4.818pt}}
\put(220,68){\makebox(0,0){6}}
\put(220.0,857.0){\rule[-0.200pt]{0.400pt}{4.818pt}}
\put(394.0,113.0){\rule[-0.200pt]{0.400pt}{4.818pt}}
\put(394,68){\makebox(0,0){$\uparrow$}}
\put(394.0,857.0){\rule[-0.200pt]{0.400pt}{4.818pt}}
\put(567.0,113.0){\rule[-0.200pt]{0.400pt}{4.818pt}}
\put(567,68){\makebox(0,0){10}}
\put(567.0,857.0){\rule[-0.200pt]{0.400pt}{4.818pt}}
\put(741.0,113.0){\rule[-0.200pt]{0.400pt}{4.818pt}}
\put(741,68){\makebox(0,0){12}}
\put(741.0,857.0){\rule[-0.200pt]{0.400pt}{4.818pt}}
\put(915.0,113.0){\rule[-0.200pt]{0.400pt}{4.818pt}}
\put(915,68){\makebox(0,0){14}}
\put(915.0,857.0){\rule[-0.200pt]{0.400pt}{4.818pt}}
\put(1089.0,113.0){\rule[-0.200pt]{0.400pt}{4.818pt}}
\put(1089,68){\makebox(0,0){16}}
\put(1089.0,857.0){\rule[-0.200pt]{0.400pt}{4.818pt}}
\put(1262.0,113.0){\rule[-0.200pt]{0.400pt}{4.818pt}}
\put(1262,68){\makebox(0,0){$\uparrow$}}
\put(1262.0,857.0){\rule[-0.200pt]{0.400pt}{4.818pt}}
\put(1436.0,113.0){\rule[-0.200pt]{0.400pt}{4.818pt}}
\put(1436,68){\makebox(0,0){20}}
\put(1436.0,857.0){\rule[-0.200pt]{0.400pt}{4.818pt}}
\put(220.0,113.0){\rule[-0.200pt]{292.934pt}{0.400pt}}
\put(1436.0,113.0){\rule[-0.200pt]{0.400pt}{184.048pt}}
\put(220.0,877.0){\rule[-0.200pt]{292.934pt}{0.400pt}}
\put(45,495){\makebox(0,0){$m^2$}}
\put(828,23){\makebox(0,0){$M^2$}}
\put(220.0,113.0){\rule[-0.200pt]{0.400pt}{184.048pt}}
\put(1306,812){\makebox(0,0)[r]{a}}
\put(1328.0,812.0){\rule[-0.200pt]{15.899pt}{0.400pt}}
\put(220,113){\usebox{\plotpoint}}
\multiput(220.58,113.00)(0.492,1.013){21}{\rule{0.119pt}{0.900pt}}
\multiput(219.17,113.00)(12.000,22.132){2}{\rule{0.400pt}{0.450pt}}
\multiput(232.58,137.00)(0.493,0.853){23}{\rule{0.119pt}{0.777pt}}
\multiput(231.17,137.00)(13.000,20.387){2}{\rule{0.400pt}{0.388pt}}
\multiput(245.58,159.00)(0.492,0.884){21}{\rule{0.119pt}{0.800pt}}
\multiput(244.17,159.00)(12.000,19.340){2}{\rule{0.400pt}{0.400pt}}
\multiput(257.58,180.00)(0.492,0.841){21}{\rule{0.119pt}{0.767pt}}
\multiput(256.17,180.00)(12.000,18.409){2}{\rule{0.400pt}{0.383pt}}
\multiput(269.58,200.00)(0.492,0.755){21}{\rule{0.119pt}{0.700pt}}
\multiput(268.17,200.00)(12.000,16.547){2}{\rule{0.400pt}{0.350pt}}
\multiput(281.58,218.00)(0.493,0.655){23}{\rule{0.119pt}{0.623pt}}
\multiput(280.17,218.00)(13.000,15.707){2}{\rule{0.400pt}{0.312pt}}
\multiput(294.58,235.00)(0.492,0.669){21}{\rule{0.119pt}{0.633pt}}
\multiput(293.17,235.00)(12.000,14.685){2}{\rule{0.400pt}{0.317pt}}
\multiput(306.58,251.00)(0.492,0.625){21}{\rule{0.119pt}{0.600pt}}
\multiput(305.17,251.00)(12.000,13.755){2}{\rule{0.400pt}{0.300pt}}
\multiput(318.58,266.00)(0.493,0.536){23}{\rule{0.119pt}{0.531pt}}
\multiput(317.17,266.00)(13.000,12.898){2}{\rule{0.400pt}{0.265pt}}
\multiput(331.58,280.00)(0.492,0.539){21}{\rule{0.119pt}{0.533pt}}
\multiput(330.17,280.00)(12.000,11.893){2}{\rule{0.400pt}{0.267pt}}
\multiput(343.00,293.58)(0.496,0.492){21}{\rule{0.500pt}{0.119pt}}
\multiput(343.00,292.17)(10.962,12.000){2}{\rule{0.250pt}{0.400pt}}
\multiput(355.00,305.58)(0.543,0.492){19}{\rule{0.536pt}{0.118pt}}
\multiput(355.00,304.17)(10.887,11.000){2}{\rule{0.268pt}{0.400pt}}
\multiput(367.00,316.58)(0.590,0.492){19}{\rule{0.573pt}{0.118pt}}
\multiput(367.00,315.17)(11.811,11.000){2}{\rule{0.286pt}{0.400pt}}
\multiput(380.00,327.58)(0.600,0.491){17}{\rule{0.580pt}{0.118pt}}
\multiput(380.00,326.17)(10.796,10.000){2}{\rule{0.290pt}{0.400pt}}
\multiput(392.00,337.58)(0.600,0.491){17}{\rule{0.580pt}{0.118pt}}
\multiput(392.00,336.17)(10.796,10.000){2}{\rule{0.290pt}{0.400pt}}
\multiput(404.00,347.59)(0.728,0.489){15}{\rule{0.678pt}{0.118pt}}
\multiput(404.00,346.17)(11.593,9.000){2}{\rule{0.339pt}{0.400pt}}
\multiput(417.00,356.59)(0.758,0.488){13}{\rule{0.700pt}{0.117pt}}
\multiput(417.00,355.17)(10.547,8.000){2}{\rule{0.350pt}{0.400pt}}
\multiput(429.00,364.59)(0.758,0.488){13}{\rule{0.700pt}{0.117pt}}
\multiput(429.00,363.17)(10.547,8.000){2}{\rule{0.350pt}{0.400pt}}
\multiput(441.00,372.59)(0.758,0.488){13}{\rule{0.700pt}{0.117pt}}
\multiput(441.00,371.17)(10.547,8.000){2}{\rule{0.350pt}{0.400pt}}
\multiput(453.00,380.59)(0.950,0.485){11}{\rule{0.843pt}{0.117pt}}
\multiput(453.00,379.17)(11.251,7.000){2}{\rule{0.421pt}{0.400pt}}
\multiput(466.00,387.59)(1.033,0.482){9}{\rule{0.900pt}{0.116pt}}
\multiput(466.00,386.17)(10.132,6.000){2}{\rule{0.450pt}{0.400pt}}
\multiput(478.00,393.59)(0.874,0.485){11}{\rule{0.786pt}{0.117pt}}
\multiput(478.00,392.17)(10.369,7.000){2}{\rule{0.393pt}{0.400pt}}
\multiput(490.00,400.59)(1.123,0.482){9}{\rule{0.967pt}{0.116pt}}
\multiput(490.00,399.17)(10.994,6.000){2}{\rule{0.483pt}{0.400pt}}
\multiput(503.00,406.59)(1.033,0.482){9}{\rule{0.900pt}{0.116pt}}
\multiput(503.00,405.17)(10.132,6.000){2}{\rule{0.450pt}{0.400pt}}
\multiput(515.00,412.59)(1.267,0.477){7}{\rule{1.060pt}{0.115pt}}
\multiput(515.00,411.17)(9.800,5.000){2}{\rule{0.530pt}{0.400pt}}
\multiput(527.00,417.59)(1.267,0.477){7}{\rule{1.060pt}{0.115pt}}
\multiput(527.00,416.17)(9.800,5.000){2}{\rule{0.530pt}{0.400pt}}
\multiput(539.00,422.59)(1.378,0.477){7}{\rule{1.140pt}{0.115pt}}
\multiput(539.00,421.17)(10.634,5.000){2}{\rule{0.570pt}{0.400pt}}
\multiput(552.00,427.59)(1.267,0.477){7}{\rule{1.060pt}{0.115pt}}
\multiput(552.00,426.17)(9.800,5.000){2}{\rule{0.530pt}{0.400pt}}
\multiput(564.00,432.60)(1.651,0.468){5}{\rule{1.300pt}{0.113pt}}
\multiput(564.00,431.17)(9.302,4.000){2}{\rule{0.650pt}{0.400pt}}
\multiput(576.00,436.59)(1.267,0.477){7}{\rule{1.060pt}{0.115pt}}
\multiput(576.00,435.17)(9.800,5.000){2}{\rule{0.530pt}{0.400pt}}
\multiput(588.00,441.60)(1.797,0.468){5}{\rule{1.400pt}{0.113pt}}
\multiput(588.00,440.17)(10.094,4.000){2}{\rule{0.700pt}{0.400pt}}
\multiput(601.00,445.60)(1.651,0.468){5}{\rule{1.300pt}{0.113pt}}
\multiput(601.00,444.17)(9.302,4.000){2}{\rule{0.650pt}{0.400pt}}
\multiput(613.00,449.61)(2.472,0.447){3}{\rule{1.700pt}{0.108pt}}
\multiput(613.00,448.17)(8.472,3.000){2}{\rule{0.850pt}{0.400pt}}
\multiput(625.00,452.60)(1.797,0.468){5}{\rule{1.400pt}{0.113pt}}
\multiput(625.00,451.17)(10.094,4.000){2}{\rule{0.700pt}{0.400pt}}
\multiput(638.00,456.61)(2.472,0.447){3}{\rule{1.700pt}{0.108pt}}
\multiput(638.00,455.17)(8.472,3.000){2}{\rule{0.850pt}{0.400pt}}
\multiput(650.00,459.60)(1.651,0.468){5}{\rule{1.300pt}{0.113pt}}
\multiput(650.00,458.17)(9.302,4.000){2}{\rule{0.650pt}{0.400pt}}
\multiput(662.00,463.61)(2.472,0.447){3}{\rule{1.700pt}{0.108pt}}
\multiput(662.00,462.17)(8.472,3.000){2}{\rule{0.850pt}{0.400pt}}
\multiput(674.00,466.61)(2.695,0.447){3}{\rule{1.833pt}{0.108pt}}
\multiput(674.00,465.17)(9.195,3.000){2}{\rule{0.917pt}{0.400pt}}
\multiput(687.00,469.61)(2.472,0.447){3}{\rule{1.700pt}{0.108pt}}
\multiput(687.00,468.17)(8.472,3.000){2}{\rule{0.850pt}{0.400pt}}
\put(699,472.17){\rule{2.500pt}{0.400pt}}
\multiput(699.00,471.17)(6.811,2.000){2}{\rule{1.250pt}{0.400pt}}
\multiput(711.00,474.61)(2.695,0.447){3}{\rule{1.833pt}{0.108pt}}
\multiput(711.00,473.17)(9.195,3.000){2}{\rule{0.917pt}{0.400pt}}
\multiput(724.00,477.61)(2.472,0.447){3}{\rule{1.700pt}{0.108pt}}
\multiput(724.00,476.17)(8.472,3.000){2}{\rule{0.850pt}{0.400pt}}
\put(736,480.17){\rule{2.500pt}{0.400pt}}
\multiput(736.00,479.17)(6.811,2.000){2}{\rule{1.250pt}{0.400pt}}
\put(748,482.17){\rule{2.500pt}{0.400pt}}
\multiput(748.00,481.17)(6.811,2.000){2}{\rule{1.250pt}{0.400pt}}
\multiput(760.00,484.61)(2.695,0.447){3}{\rule{1.833pt}{0.108pt}}
\multiput(760.00,483.17)(9.195,3.000){2}{\rule{0.917pt}{0.400pt}}
\put(773,487.17){\rule{2.500pt}{0.400pt}}
\multiput(773.00,486.17)(6.811,2.000){2}{\rule{1.250pt}{0.400pt}}
\put(785,489.17){\rule{2.500pt}{0.400pt}}
\multiput(785.00,488.17)(6.811,2.000){2}{\rule{1.250pt}{0.400pt}}
\put(797,491.17){\rule{2.700pt}{0.400pt}}
\multiput(797.00,490.17)(7.396,2.000){2}{\rule{1.350pt}{0.400pt}}
\put(810,493.17){\rule{2.500pt}{0.400pt}}
\multiput(810.00,492.17)(6.811,2.000){2}{\rule{1.250pt}{0.400pt}}
\put(822,495.17){\rule{2.500pt}{0.400pt}}
\multiput(822.00,494.17)(6.811,2.000){2}{\rule{1.250pt}{0.400pt}}
\put(834,497.17){\rule{2.500pt}{0.400pt}}
\multiput(834.00,496.17)(6.811,2.000){2}{\rule{1.250pt}{0.400pt}}
\put(846,499.17){\rule{2.700pt}{0.400pt}}
\multiput(846.00,498.17)(7.396,2.000){2}{\rule{1.350pt}{0.400pt}}
\put(859,500.67){\rule{2.891pt}{0.400pt}}
\multiput(859.00,500.17)(6.000,1.000){2}{\rule{1.445pt}{0.400pt}}
\put(871,502.17){\rule{2.500pt}{0.400pt}}
\multiput(871.00,501.17)(6.811,2.000){2}{\rule{1.250pt}{0.400pt}}
\put(883,504.17){\rule{2.700pt}{0.400pt}}
\multiput(883.00,503.17)(7.396,2.000){2}{\rule{1.350pt}{0.400pt}}
\put(896,505.67){\rule{2.891pt}{0.400pt}}
\multiput(896.00,505.17)(6.000,1.000){2}{\rule{1.445pt}{0.400pt}}
\put(908,507.17){\rule{2.500pt}{0.400pt}}
\multiput(908.00,506.17)(6.811,2.000){2}{\rule{1.250pt}{0.400pt}}
\put(920,508.67){\rule{2.891pt}{0.400pt}}
\multiput(920.00,508.17)(6.000,1.000){2}{\rule{1.445pt}{0.400pt}}
\put(932,509.67){\rule{3.132pt}{0.400pt}}
\multiput(932.00,509.17)(6.500,1.000){2}{\rule{1.566pt}{0.400pt}}
\put(945,511.17){\rule{2.500pt}{0.400pt}}
\multiput(945.00,510.17)(6.811,2.000){2}{\rule{1.250pt}{0.400pt}}
\put(957,512.67){\rule{2.891pt}{0.400pt}}
\multiput(957.00,512.17)(6.000,1.000){2}{\rule{1.445pt}{0.400pt}}
\put(969,513.67){\rule{3.132pt}{0.400pt}}
\multiput(969.00,513.17)(6.500,1.000){2}{\rule{1.566pt}{0.400pt}}
\put(982,515.17){\rule{2.500pt}{0.400pt}}
\multiput(982.00,514.17)(6.811,2.000){2}{\rule{1.250pt}{0.400pt}}
\put(994,516.67){\rule{2.891pt}{0.400pt}}
\multiput(994.00,516.17)(6.000,1.000){2}{\rule{1.445pt}{0.400pt}}
\put(1006,517.67){\rule{2.891pt}{0.400pt}}
\multiput(1006.00,517.17)(6.000,1.000){2}{\rule{1.445pt}{0.400pt}}
\put(1018,518.67){\rule{3.132pt}{0.400pt}}
\multiput(1018.00,518.17)(6.500,1.000){2}{\rule{1.566pt}{0.400pt}}
\put(1031,519.67){\rule{2.891pt}{0.400pt}}
\multiput(1031.00,519.17)(6.000,1.000){2}{\rule{1.445pt}{0.400pt}}
\put(1043,520.67){\rule{2.891pt}{0.400pt}}
\multiput(1043.00,520.17)(6.000,1.000){2}{\rule{1.445pt}{0.400pt}}
\put(1055,521.67){\rule{3.132pt}{0.400pt}}
\multiput(1055.00,521.17)(6.500,1.000){2}{\rule{1.566pt}{0.400pt}}
\put(1068,523.17){\rule{2.500pt}{0.400pt}}
\multiput(1068.00,522.17)(6.811,2.000){2}{\rule{1.250pt}{0.400pt}}
\put(1080,524.67){\rule{2.891pt}{0.400pt}}
\multiput(1080.00,524.17)(6.000,1.000){2}{\rule{1.445pt}{0.400pt}}
\put(1104,525.67){\rule{3.132pt}{0.400pt}}
\multiput(1104.00,525.17)(6.500,1.000){2}{\rule{1.566pt}{0.400pt}}
\put(1117,526.67){\rule{2.891pt}{0.400pt}}
\multiput(1117.00,526.17)(6.000,1.000){2}{\rule{1.445pt}{0.400pt}}
\put(1129,527.67){\rule{2.891pt}{0.400pt}}
\multiput(1129.00,527.17)(6.000,1.000){2}{\rule{1.445pt}{0.400pt}}
\put(1141,528.67){\rule{2.891pt}{0.400pt}}
\multiput(1141.00,528.17)(6.000,1.000){2}{\rule{1.445pt}{0.400pt}}
\put(1153,529.67){\rule{3.132pt}{0.400pt}}
\multiput(1153.00,529.17)(6.500,1.000){2}{\rule{1.566pt}{0.400pt}}
\put(1166,530.67){\rule{2.891pt}{0.400pt}}
\multiput(1166.00,530.17)(6.000,1.000){2}{\rule{1.445pt}{0.400pt}}
\put(1178,531.67){\rule{2.891pt}{0.400pt}}
\multiput(1178.00,531.17)(6.000,1.000){2}{\rule{1.445pt}{0.400pt}}
\put(1092.0,526.0){\rule[-0.200pt]{2.891pt}{0.400pt}}
\put(1203,532.67){\rule{2.891pt}{0.400pt}}
\multiput(1203.00,532.17)(6.000,1.000){2}{\rule{1.445pt}{0.400pt}}
\put(1215,533.67){\rule{2.891pt}{0.400pt}}
\multiput(1215.00,533.17)(6.000,1.000){2}{\rule{1.445pt}{0.400pt}}
\put(1227,534.67){\rule{2.891pt}{0.400pt}}
\multiput(1227.00,534.17)(6.000,1.000){2}{\rule{1.445pt}{0.400pt}}
\put(1190.0,533.0){\rule[-0.200pt]{3.132pt}{0.400pt}}
\put(1252,535.67){\rule{2.891pt}{0.400pt}}
\multiput(1252.00,535.17)(6.000,1.000){2}{\rule{1.445pt}{0.400pt}}
\put(1264,536.67){\rule{2.891pt}{0.400pt}}
\multiput(1264.00,536.17)(6.000,1.000){2}{\rule{1.445pt}{0.400pt}}
\put(1239.0,536.0){\rule[-0.200pt]{3.132pt}{0.400pt}}
\put(1289,537.67){\rule{2.891pt}{0.400pt}}
\multiput(1289.00,537.17)(6.000,1.000){2}{\rule{1.445pt}{0.400pt}}
\put(1301,538.67){\rule{2.891pt}{0.400pt}}
\multiput(1301.00,538.17)(6.000,1.000){2}{\rule{1.445pt}{0.400pt}}
\put(1276.0,538.0){\rule[-0.200pt]{3.132pt}{0.400pt}}
\put(1325,539.67){\rule{3.132pt}{0.400pt}}
\multiput(1325.00,539.17)(6.500,1.000){2}{\rule{1.566pt}{0.400pt}}
\put(1338,540.67){\rule{2.891pt}{0.400pt}}
\multiput(1338.00,540.17)(6.000,1.000){2}{\rule{1.445pt}{0.400pt}}
\put(1313.0,540.0){\rule[-0.200pt]{2.891pt}{0.400pt}}
\put(1362,541.67){\rule{3.132pt}{0.400pt}}
\multiput(1362.00,541.17)(6.500,1.000){2}{\rule{1.566pt}{0.400pt}}
\put(1350.0,542.0){\rule[-0.200pt]{2.891pt}{0.400pt}}
\put(1387,542.67){\rule{2.891pt}{0.400pt}}
\multiput(1387.00,542.17)(6.000,1.000){2}{\rule{1.445pt}{0.400pt}}
\put(1375.0,543.0){\rule[-0.200pt]{2.891pt}{0.400pt}}
\put(1411,543.67){\rule{3.132pt}{0.400pt}}
\multiput(1411.00,543.17)(6.500,1.000){2}{\rule{1.566pt}{0.400pt}}
\put(1424,544.67){\rule{2.891pt}{0.400pt}}
\multiput(1424.00,544.17)(6.000,1.000){2}{\rule{1.445pt}{0.400pt}}
\put(1399.0,544.0){\rule[-0.200pt]{2.891pt}{0.400pt}}
\sbox{\plotpoint}{\rule[-0.500pt]{1.000pt}{1.000pt}}%
\put(1306,767){\makebox(0,0)[r]{b}}
\multiput(1328,767)(20.756,0.000){4}{\usebox{\plotpoint}}
\put(1394,767){\usebox{\plotpoint}}
\put(220,877){\usebox{\plotpoint}}
\multiput(220,877)(8.982,-18.712){2}{\usebox{\plotpoint}}
\put(239.01,840.14){\usebox{\plotpoint}}
\put(249.98,822.53){\usebox{\plotpoint}}
\put(261.67,805.38){\usebox{\plotpoint}}
\put(274.03,788.71){\usebox{\plotpoint}}
\put(287.79,773.21){\usebox{\plotpoint}}
\put(302.82,758.91){\usebox{\plotpoint}}
\multiput(306,756)(15.300,-14.025){0}{\usebox{\plotpoint}}
\put(318.14,744.91){\usebox{\plotpoint}}
\put(335.09,732.93){\usebox{\plotpoint}}
\put(352.04,720.97){\usebox{\plotpoint}}
\multiput(355,719)(17.928,-10.458){0}{\usebox{\plotpoint}}
\put(370.00,710.61){\usebox{\plotpoint}}
\put(388.72,701.64){\usebox{\plotpoint}}
\multiput(392,700)(18.564,-9.282){0}{\usebox{\plotpoint}}
\put(407.42,692.68){\usebox{\plotpoint}}
\put(426.69,684.96){\usebox{\plotpoint}}
\multiput(429,684)(19.690,-6.563){0}{\usebox{\plotpoint}}
\put(446.31,678.23){\usebox{\plotpoint}}
\multiput(453,676)(19.838,-6.104){0}{\usebox{\plotpoint}}
\put(466.10,671.97){\usebox{\plotpoint}}
\put(486.06,666.31){\usebox{\plotpoint}}
\multiput(490,665)(20.224,-4.667){0}{\usebox{\plotpoint}}
\put(506.21,661.46){\usebox{\plotpoint}}
\put(526.49,657.13){\usebox{\plotpoint}}
\multiput(527,657)(20.136,-5.034){0}{\usebox{\plotpoint}}
\put(546.77,652.80){\usebox{\plotpoint}}
\multiput(552,652)(20.473,-3.412){0}{\usebox{\plotpoint}}
\put(567.25,649.46){\usebox{\plotpoint}}
\put(587.73,646.05){\usebox{\plotpoint}}
\multiput(588,646)(20.514,-3.156){0}{\usebox{\plotpoint}}
\put(608.23,642.80){\usebox{\plotpoint}}
\multiput(613,642)(20.473,-3.412){0}{\usebox{\plotpoint}}
\put(628.74,639.71){\usebox{\plotpoint}}
\put(649.31,637.11){\usebox{\plotpoint}}
\multiput(650,637)(20.684,-1.724){0}{\usebox{\plotpoint}}
\put(669.99,635.33){\usebox{\plotpoint}}
\multiput(674,635)(20.514,-3.156){0}{\usebox{\plotpoint}}
\put(690.57,632.70){\usebox{\plotpoint}}
\multiput(699,632)(20.684,-1.724){0}{\usebox{\plotpoint}}
\put(711.25,630.98){\usebox{\plotpoint}}
\put(731.94,629.34){\usebox{\plotpoint}}
\multiput(736,629)(20.684,-1.724){0}{\usebox{\plotpoint}}
\put(752.62,627.61){\usebox{\plotpoint}}
\multiput(760,627)(20.694,-1.592){0}{\usebox{\plotpoint}}
\put(773.31,625.97){\usebox{\plotpoint}}
\put(794.00,624.25){\usebox{\plotpoint}}
\multiput(797,624)(20.694,-1.592){0}{\usebox{\plotpoint}}
\put(814.70,623.00){\usebox{\plotpoint}}
\multiput(822,623)(20.684,-1.724){0}{\usebox{\plotpoint}}
\put(835.41,621.88){\usebox{\plotpoint}}
\put(856.10,620.22){\usebox{\plotpoint}}
\multiput(859,620)(20.756,0.000){0}{\usebox{\plotpoint}}
\put(876.83,619.51){\usebox{\plotpoint}}
\multiput(883,619)(20.694,-1.592){0}{\usebox{\plotpoint}}
\put(897.52,618.00){\usebox{\plotpoint}}
\put(918.24,617.15){\usebox{\plotpoint}}
\multiput(920,617)(20.756,0.000){0}{\usebox{\plotpoint}}
\put(938.97,616.46){\usebox{\plotpoint}}
\multiput(945,616)(20.756,0.000){0}{\usebox{\plotpoint}}
\put(959.70,615.77){\usebox{\plotpoint}}
\put(980.42,615.00){\usebox{\plotpoint}}
\multiput(982,615)(20.684,-1.724){0}{\usebox{\plotpoint}}
\put(1001.14,614.00){\usebox{\plotpoint}}
\multiput(1006,614)(20.684,-1.724){0}{\usebox{\plotpoint}}
\put(1021.85,613.00){\usebox{\plotpoint}}
\put(1042.61,613.00){\usebox{\plotpoint}}
\multiput(1043,613)(20.684,-1.724){0}{\usebox{\plotpoint}}
\put(1063.32,612.00){\usebox{\plotpoint}}
\multiput(1068,612)(20.756,0.000){0}{\usebox{\plotpoint}}
\put(1084.06,611.66){\usebox{\plotpoint}}
\multiput(1092,611)(20.756,0.000){0}{\usebox{\plotpoint}}
\put(1104.79,610.94){\usebox{\plotpoint}}
\put(1125.51,610.00){\usebox{\plotpoint}}
\multiput(1129,610)(20.756,0.000){0}{\usebox{\plotpoint}}
\put(1146.26,610.00){\usebox{\plotpoint}}
\multiput(1153,610)(20.694,-1.592){0}{\usebox{\plotpoint}}
\put(1166.98,609.00){\usebox{\plotpoint}}
\put(1187.74,609.00){\usebox{\plotpoint}}
\multiput(1190,609)(20.694,-1.592){0}{\usebox{\plotpoint}}
\put(1208.45,608.00){\usebox{\plotpoint}}
\multiput(1215,608)(20.756,0.000){0}{\usebox{\plotpoint}}
\put(1229.21,608.00){\usebox{\plotpoint}}
\put(1249.93,607.16){\usebox{\plotpoint}}
\multiput(1252,607)(20.756,0.000){0}{\usebox{\plotpoint}}
\put(1270.68,607.00){\usebox{\plotpoint}}
\multiput(1276,607)(20.756,0.000){0}{\usebox{\plotpoint}}
\put(1291.44,607.00){\usebox{\plotpoint}}
\put(1312.15,606.07){\usebox{\plotpoint}}
\multiput(1313,606)(20.756,0.000){0}{\usebox{\plotpoint}}
\put(1332.91,606.00){\usebox{\plotpoint}}
\multiput(1338,606)(20.756,0.000){0}{\usebox{\plotpoint}}
\put(1353.66,606.00){\usebox{\plotpoint}}
\put(1374.38,605.05){\usebox{\plotpoint}}
\multiput(1375,605)(20.756,0.000){0}{\usebox{\plotpoint}}
\put(1395.13,605.00){\usebox{\plotpoint}}
\multiput(1399,605)(20.756,0.000){0}{\usebox{\plotpoint}}
\put(1415.89,605.00){\usebox{\plotpoint}}
\multiput(1424,605)(20.756,0.000){0}{\usebox{\plotpoint}}
\put(1436,605){\usebox{\plotpoint}}
\sbox{\plotpoint}{\rule[-0.400pt]{0.800pt}{0.800pt}}%
\put(1306,722){\makebox(0,0)[r]{c}}
\put(1328.0,722.0){\rule[-0.400pt]{15.899pt}{0.800pt}}
\put(220,674){\usebox{\plotpoint}}
\put(220,670.84){\rule{2.891pt}{0.800pt}}
\multiput(220.00,672.34)(6.000,-3.000){2}{\rule{1.445pt}{0.800pt}}
\put(232,667.34){\rule{2.800pt}{0.800pt}}
\multiput(232.00,669.34)(7.188,-4.000){2}{\rule{1.400pt}{0.800pt}}
\put(245,663.84){\rule{2.891pt}{0.800pt}}
\multiput(245.00,665.34)(6.000,-3.000){2}{\rule{1.445pt}{0.800pt}}
\put(257,660.84){\rule{2.891pt}{0.800pt}}
\multiput(257.00,662.34)(6.000,-3.000){2}{\rule{1.445pt}{0.800pt}}
\put(269,657.84){\rule{2.891pt}{0.800pt}}
\multiput(269.00,659.34)(6.000,-3.000){2}{\rule{1.445pt}{0.800pt}}
\put(281,655.34){\rule{3.132pt}{0.800pt}}
\multiput(281.00,656.34)(6.500,-2.000){2}{\rule{1.566pt}{0.800pt}}
\put(294,652.84){\rule{2.891pt}{0.800pt}}
\multiput(294.00,654.34)(6.000,-3.000){2}{\rule{1.445pt}{0.800pt}}
\put(306,650.34){\rule{2.891pt}{0.800pt}}
\multiput(306.00,651.34)(6.000,-2.000){2}{\rule{1.445pt}{0.800pt}}
\put(318,648.34){\rule{3.132pt}{0.800pt}}
\multiput(318.00,649.34)(6.500,-2.000){2}{\rule{1.566pt}{0.800pt}}
\put(331,645.84){\rule{2.891pt}{0.800pt}}
\multiput(331.00,647.34)(6.000,-3.000){2}{\rule{1.445pt}{0.800pt}}
\put(343,643.34){\rule{2.891pt}{0.800pt}}
\multiput(343.00,644.34)(6.000,-2.000){2}{\rule{1.445pt}{0.800pt}}
\put(355,641.84){\rule{2.891pt}{0.800pt}}
\multiput(355.00,642.34)(6.000,-1.000){2}{\rule{1.445pt}{0.800pt}}
\put(367,640.34){\rule{3.132pt}{0.800pt}}
\multiput(367.00,641.34)(6.500,-2.000){2}{\rule{1.566pt}{0.800pt}}
\put(380,638.34){\rule{2.891pt}{0.800pt}}
\multiput(380.00,639.34)(6.000,-2.000){2}{\rule{1.445pt}{0.800pt}}
\put(392,636.34){\rule{2.891pt}{0.800pt}}
\multiput(392.00,637.34)(6.000,-2.000){2}{\rule{1.445pt}{0.800pt}}
\put(404,634.84){\rule{3.132pt}{0.800pt}}
\multiput(404.00,635.34)(6.500,-1.000){2}{\rule{1.566pt}{0.800pt}}
\put(417,633.34){\rule{2.891pt}{0.800pt}}
\multiput(417.00,634.34)(6.000,-2.000){2}{\rule{1.445pt}{0.800pt}}
\put(429,631.84){\rule{2.891pt}{0.800pt}}
\multiput(429.00,632.34)(6.000,-1.000){2}{\rule{1.445pt}{0.800pt}}
\put(441,630.84){\rule{2.891pt}{0.800pt}}
\multiput(441.00,631.34)(6.000,-1.000){2}{\rule{1.445pt}{0.800pt}}
\put(453,629.34){\rule{3.132pt}{0.800pt}}
\multiput(453.00,630.34)(6.500,-2.000){2}{\rule{1.566pt}{0.800pt}}
\put(466,627.84){\rule{2.891pt}{0.800pt}}
\multiput(466.00,628.34)(6.000,-1.000){2}{\rule{1.445pt}{0.800pt}}
\put(478,626.84){\rule{2.891pt}{0.800pt}}
\multiput(478.00,627.34)(6.000,-1.000){2}{\rule{1.445pt}{0.800pt}}
\put(490,625.84){\rule{3.132pt}{0.800pt}}
\multiput(490.00,626.34)(6.500,-1.000){2}{\rule{1.566pt}{0.800pt}}
\put(503,624.84){\rule{2.891pt}{0.800pt}}
\multiput(503.00,625.34)(6.000,-1.000){2}{\rule{1.445pt}{0.800pt}}
\put(515,623.84){\rule{2.891pt}{0.800pt}}
\multiput(515.00,624.34)(6.000,-1.000){2}{\rule{1.445pt}{0.800pt}}
\put(527,622.84){\rule{2.891pt}{0.800pt}}
\multiput(527.00,623.34)(6.000,-1.000){2}{\rule{1.445pt}{0.800pt}}
\put(539,621.84){\rule{3.132pt}{0.800pt}}
\multiput(539.00,622.34)(6.500,-1.000){2}{\rule{1.566pt}{0.800pt}}
\put(552,620.84){\rule{2.891pt}{0.800pt}}
\multiput(552.00,621.34)(6.000,-1.000){2}{\rule{1.445pt}{0.800pt}}
\put(564,619.84){\rule{2.891pt}{0.800pt}}
\multiput(564.00,620.34)(6.000,-1.000){2}{\rule{1.445pt}{0.800pt}}
\put(588,618.84){\rule{3.132pt}{0.800pt}}
\multiput(588.00,619.34)(6.500,-1.000){2}{\rule{1.566pt}{0.800pt}}
\put(601,617.84){\rule{2.891pt}{0.800pt}}
\multiput(601.00,618.34)(6.000,-1.000){2}{\rule{1.445pt}{0.800pt}}
\put(613,616.84){\rule{2.891pt}{0.800pt}}
\multiput(613.00,617.34)(6.000,-1.000){2}{\rule{1.445pt}{0.800pt}}
\put(576.0,621.0){\rule[-0.400pt]{2.891pt}{0.800pt}}
\put(638,615.84){\rule{2.891pt}{0.800pt}}
\multiput(638.00,616.34)(6.000,-1.000){2}{\rule{1.445pt}{0.800pt}}
\put(625.0,618.0){\rule[-0.400pt]{3.132pt}{0.800pt}}
\put(662,614.84){\rule{2.891pt}{0.800pt}}
\multiput(662.00,615.34)(6.000,-1.000){2}{\rule{1.445pt}{0.800pt}}
\put(674,613.84){\rule{3.132pt}{0.800pt}}
\multiput(674.00,614.34)(6.500,-1.000){2}{\rule{1.566pt}{0.800pt}}
\put(650.0,617.0){\rule[-0.400pt]{2.891pt}{0.800pt}}
\put(699,612.84){\rule{2.891pt}{0.800pt}}
\multiput(699.00,613.34)(6.000,-1.000){2}{\rule{1.445pt}{0.800pt}}
\put(687.0,615.0){\rule[-0.400pt]{2.891pt}{0.800pt}}
\put(724,611.84){\rule{2.891pt}{0.800pt}}
\multiput(724.00,612.34)(6.000,-1.000){2}{\rule{1.445pt}{0.800pt}}
\put(711.0,614.0){\rule[-0.400pt]{3.132pt}{0.800pt}}
\put(748,610.84){\rule{2.891pt}{0.800pt}}
\multiput(748.00,611.34)(6.000,-1.000){2}{\rule{1.445pt}{0.800pt}}
\put(736.0,613.0){\rule[-0.400pt]{2.891pt}{0.800pt}}
\put(785,609.84){\rule{2.891pt}{0.800pt}}
\multiput(785.00,610.34)(6.000,-1.000){2}{\rule{1.445pt}{0.800pt}}
\put(760.0,612.0){\rule[-0.400pt]{6.022pt}{0.800pt}}
\put(810,608.84){\rule{2.891pt}{0.800pt}}
\multiput(810.00,609.34)(6.000,-1.000){2}{\rule{1.445pt}{0.800pt}}
\put(797.0,611.0){\rule[-0.400pt]{3.132pt}{0.800pt}}
\put(846,607.84){\rule{3.132pt}{0.800pt}}
\multiput(846.00,608.34)(6.500,-1.000){2}{\rule{1.566pt}{0.800pt}}
\put(822.0,610.0){\rule[-0.400pt]{5.782pt}{0.800pt}}
\put(883,606.84){\rule{3.132pt}{0.800pt}}
\multiput(883.00,607.34)(6.500,-1.000){2}{\rule{1.566pt}{0.800pt}}
\put(859.0,609.0){\rule[-0.400pt]{5.782pt}{0.800pt}}
\put(932,605.84){\rule{3.132pt}{0.800pt}}
\multiput(932.00,606.34)(6.500,-1.000){2}{\rule{1.566pt}{0.800pt}}
\put(896.0,608.0){\rule[-0.400pt]{8.672pt}{0.800pt}}
\put(982,604.84){\rule{2.891pt}{0.800pt}}
\multiput(982.00,605.34)(6.000,-1.000){2}{\rule{1.445pt}{0.800pt}}
\put(945.0,607.0){\rule[-0.400pt]{8.913pt}{0.800pt}}
\put(1031,603.84){\rule{2.891pt}{0.800pt}}
\multiput(1031.00,604.34)(6.000,-1.000){2}{\rule{1.445pt}{0.800pt}}
\put(994.0,606.0){\rule[-0.400pt]{8.913pt}{0.800pt}}
\put(1104,602.84){\rule{3.132pt}{0.800pt}}
\multiput(1104.00,603.34)(6.500,-1.000){2}{\rule{1.566pt}{0.800pt}}
\put(1043.0,605.0){\rule[-0.400pt]{14.695pt}{0.800pt}}
\put(1178,601.84){\rule{2.891pt}{0.800pt}}
\multiput(1178.00,602.34)(6.000,-1.000){2}{\rule{1.445pt}{0.800pt}}
\put(1117.0,604.0){\rule[-0.400pt]{14.695pt}{0.800pt}}
\put(1264,600.84){\rule{2.891pt}{0.800pt}}
\multiput(1264.00,601.34)(6.000,-1.000){2}{\rule{1.445pt}{0.800pt}}
\put(1190.0,603.0){\rule[-0.400pt]{17.827pt}{0.800pt}}
\put(1362,599.84){\rule{3.132pt}{0.800pt}}
\multiput(1362.00,600.34)(6.500,-1.000){2}{\rule{1.566pt}{0.800pt}}
\put(1276.0,602.0){\rule[-0.400pt]{20.717pt}{0.800pt}}
\put(1375.0,601.0){\rule[-0.400pt]{14.695pt}{0.800pt}}
\end{picture}

\begin{center}
{\large \bf Fig. 2}
\end{center}


\begin{thebibliography}{99}

\bibitem {1} M.A.Shifman, A.I.Vainshtein and V.I.Zakharov,
             Nucl.Phys. {\bf B147}(1979)385.

\bibitem {r} L.J.Reinders, H.R.Rubinstein and S.Yazaky,
             Phys.Rep.  {\bf 127}(1985)1.

\bibitem {s} M.A.Shifman, "QCD sum rules: physical picture and historical
             survey" (Lectures given at 11th Autumn School of Physics, Lisbon,
             Portugal, Oct 9-14, 1989), Published in  Lisbon School 1989.

\bibitem {n} S.Narison, "QCD Spectral Sum Rules"  (Lecture
             Notes in Physics, Vol. 26), World Scientific,
             Singapore 1989.

\bibitem {3} V.A.Novikov, M.A. Shifman, A.I. Vainshtein
             and V.I. Zakharov, Nucl.Phys. {\bf B191}(1981)301.

\bibitem {4} I.I.Bigi, M.A. Shifman, N.G. Uraltsev
             and A.I. Vainshtein, preprint CERN-TH.7171/94.

\bibitem {sm} A.A.Pivovarov, N.N.Tavkhelidze and V.F.Tokarev,
              Phys.Lett. {\bf B132}(1983)402; \\
              A.A.Pivovarov and V.F.Tokarev, Yad.Fiz. {\bf 41}(1985)524.

\bibitem {5} N.I.Ussukina and A.I.Davydychev, Phys.Lett. {\bf B305}(1993)136.

\bibitem {6} V.V.Belokurov and N.I.Ussukina, J.Phys. {\bf A16}(1983)2811.

\bibitem {7} D.J.Brodhurst, Phys.Lett. {\bf B307}(1993)132.

\bibitem {8} N.V.Krasnikov and A.A.Pivovarov,
             Phys.Lett {\bf B112}(1982)397;\\
             N.V.Krasnikov, A.A.Pivovarov and N.N.Tavkhelidze,
             Z.Phys. {\bf C19}(1983)301.



\end{thebibliography}
\end{document}